\documentclass[aps,prl]{revtex4}

\usepackage{graphicx}
\usepackage{amssymb}

\newcommand \be{\begin{eqnarray}}
\newcommand \ee{\end{eqnarray}}

\begin{document}
 
\title{Stochastic Control of Metabolic Pathways}

\author{Andrea Rocco}

\affiliation{Department of Mathematical Sciences, University of Bath\\ 
BA2 7AY Bath, United Kingdom
}

\date{\today}

\begin{abstract}
We study the effect of extrinsic noise in metabolic networks. 
We introduce external random fluctuations at the kinetic level, 
and show how these lead to a stochastic generalization of 
standard Metabolic Control Analysis. While Summation and Connectivity Theorems 
hold true in presence of extrinsic noise, control coefficients 
incorporate its effect through an explicit dependency on the noise
intensity. New elasticities and response coefficients are also defined.
Accordingly, the concept of control by noise 
is introduced as a way of tuning the systemic behaviour of metabolisms. 
We argue that this framework holds for intrinsic noise too, 
when time-scale separation is present in the system. 
\end{abstract}


\maketitle

\section{1. Introduction}

Stochastic fluctuations represent an important contribution to complex
behaviours of biological systems.
Stochasticity appears as a fundamental dynamical mechanism, which does not
only generate phenotypic diversity \cite{Blake03}, but also plays a 
major role at many different levels. Selection of alternative pathways 
in epigenetic switches
\cite{Aurell02}, or synchronisation of multicellular systems 
\cite{Zhou05}, are just two of the many 
processes which not only are influenced by stochasticity, 
but seem to use it to perform in optimal way
\cite{Vilar02}. 

In molecular biology two classes of stochastic fluctuations are
particularly relevant. 
The first class is related to the low copy number of chemical species.
In particular, if $N$ is the number of molecules in the system,
fluctuations in $N$ lead to an associated statistical noise with 
intensity of the order of $N^{-1/2}$. While
continuous deterministic descriptions in terms of average concentrations can
effectively capture the relevant dynamics for $N$ large, when $N$ 
is small fluctuations may become huge, and noise cannot be neglected
\cite{Elf03}. Gene regulation is a typical example, as it may be affected
by large fluctuations due to the low copy number of transcription 
factors \cite{Becskei05}. Fluctuations associated with the
intrinsic discreteness of the collisional processes among single
molecules are usually referred to as intrinsic (or internal) noise.

On the other hand, the behaviour of a biochemical system 
depends also on a number of control parameters. Some of these relate
directly to the macroscopic environment, such as for instance illumination
conditions, or pH levels. Others have a more microscopic origin, but
still exhert a control on the sytem which is independent of
its intrinsic dynamics. In gene regulatory
networks, for instance, factors acting globally on all genes, such
as abundance of RNA polymerase, can change the global efficiency of
transcription factors, and ultimately contribute to tuning gene expression levels
\cite{Pedraza05}. Similarly, reaction constants, 
enzyme activities, or input signals critically control the functional
behaviour of metabolic pathways. In thise sense, all these parameters are external to
the system. Fluctuations of external parameters define a second type
of noise, usually referred to as extrinsic, or external.

The effect of extrinsic noise can be highly non-trivial and 
counterintuitive. Extrinsic noise in homogeneous chemical 
systems has been shown to 
provoke noise-induced transitions \cite{Lefever84}, and recently
it has been identified as a mechanism for creating and sustaining
spatio-temporal patterns in spatially extended systems 
\cite{GarciaOjalvo99}. In gene networks the interplay between intrinsic and 
extrinsic noise has been recently analysed in
\cite{Swain02,Paulsson04,TanaseNicola06}.

In this paper we plan to extend these findings to 
metabolic networks. We shall focus 
on the case when the copy number of molecules
is large enough for intrinsic noise to be negligible, so that a
continuous description in terms of average concentrations is
feasible, but at the same time the system 
experiences extrinsic stochastic fluctuations. 
Deterministic rate equations become thus stochastic differential
equations.

Given the description in terms of rate equations, it is ideally 
desirable to define other, more ``systemic'' approaches, which explain how
global properties of the pathway, such as fluxes and concentrations,
depend on local variables, such as enzyme activities. To this aim, in the specific
case of metabolic networks a useful strategy --the so-called
Metabolic Control Analysis (MCA)
\cite{Heinrich96}-- has been developed, and has become
nowadays a popular quantitative framework for investigating control and
regulation of metabolisms. 

Mathematically MCA is a sensitivity analysis, and is based on a proper 
manipulation of the rate equations. It consists in perturbing the
parameters of the system, and evaluating the corresponding change in
steady state fluxes and concentrations. 
Perturbations need to be small so that a linear
approximation can hold, even if in principle the analysis can be carried on
including higher order terms. The effect of the perturbation 
is represented by a control coefficient, which expresses
the control that a certain rate exerts on the steady state variable 
of interest. 

A fundamental property of control coefficients is their
interdependence. This is formalized in the so-called Summation Theorems  
\cite{Heinrich96} for both flux and concentration control
coefficients: For any flux, the sum of the flux control
coefficients of all the rates in the pathway equals one. Similarly, for
any metabolite, the sum of the concentration control coefficients of
all the rates in the pathway equals zero. These properties 
demonstrate the actual complexity of the system. 
In fact, control on systemic variables is typically 
distributed among all the enzymes in the system, being the respective
control coefficients often of the same order of magnitude and
constrained by the Summation Theorems. This implies
that none of them alone is capable of determining the systemic 
properties of the network \cite{Westerhoff05}. 

Control coefficients are also responsible for mediating the way local
perturbations climb up to the systemic level. This is the content 
of the so-called Connectivity Theorems and Partitioned Response
relations. The response of the system is the sum of the product of 
the local sensitivity of the individual rates to perturbations 
(elasticities) times the control that those rates exert on the system
variable of interest. Connectivity Theorems and Partitioned Response, 
together with the Summation Theorems, define the mathematical structure
of Metabolic Control Analysis. 

We present here a novel framework, which allows the description of extrinsic
fluctuations within Metabolic Control Analysis. Our approach shows that
when considering extrinsic stochasticity, Summation Theorems are still valid
in the standard form, but with control coefficients explicitly
dependent on the noise intensity. The same is true for Connectivity
Theorems and Partitioned Response, and for elasticities and response
coefficients. Our findings suggest a
reinterpretation of noise as an 
``active'' player in the global dynamics of the network, rather than as 
a mere uncontrollable variable. On the one hand, noise may allow 
the metabolic pathway to access dynamical regions which would be
otherwise out of reach if the dynamics were purely deterministic. On
the other hand, it may provide us with an alternative mechanism to
exert external control on the regulation of metabolisms.

In Section 2 we define the stochastic rate equations
associated with extrinsic noise, and we illustrate how the 
corresponding stationary probability distribution shows non-trivial 
modifications with respect to the deterministic case. 
Section 3 is the core of the
paper. When extrinsic stochasticity is implemented, the general
principles of MCA still hold, and incorporate in a natural way the
effect of noise. An explicit illustration of our approach is presented in
Section 4, and we finally conclude in Section 5.

\section{2. Kinetics with extrinsic stochasticity}\label{SKEN}

Consider a metabolic network with $N$ metabolites and $M$ reactions. 
In the homogeneous (well-stirred) situation, spatial dependencies
are negligible, and the network is dynamically represented in terms of the
following set of ordinary differential equations:
\be
\frac{d \mathbf{c}}{dt} = \mathbf{f}(\mathbf{c},
     {\boldmath{\mbox{$\mu$}}}), \label{determ}
\ee
with 
\be
f_i = \sum_{j=1}^M S_{ij} v_j, \quad i=1,...,N. \label{kin}
\ee
The $N \times M$ matrix $S_{ij}$ is the so-called stoichiometry
matrix, and $v_j = v_j(\mathbf{c},{\boldmath{\mbox{$\mu$}}})$ are
reaction rates.
Here $\mathbf{c} = \{c_1,...,c_N\}$ represents the vector of concentrations, 
and ${\boldmath{\mbox{$\mu$}}} = \{\mu_1,...,\mu_R   \}$ is a vector
of $R$ parameters, such as reaction constants, enzyme concentrations,
illumination conditions, input signals, etc.

We are interested in the situation when these parameters behave
dynamically, and in particular exhibit random fluctuations. 
Experiments can be designed 
on purpose to investigate the effect of extrinsic fluctuations. 
For instance, in \cite{Kar03} the effect of
stochastic fluctuations on the substrate injection rate in a model for
glycolytic oscillations is analyzed, and a new noise-dependent
dynamical regime of oscillations is identified.
On the other hand, extrinsic noise can originate from intrinsic noise 
after some reduction procedure has been carried out. For
instance the derivation of the Michaelis-Menten kinetics relies on 
the assumption of rapid equilibration of the enzyme 
\cite{Heinrich96,Klipp05}. As a consequence, 
the dependency on the enzyme concentration is parametric in the
Michaelis-Menten form, and intrinsic fluctuations on enzyme activity
are then perceived as extrinsic in the reduced model. 
Enzyme activity can fluctuate because of the
stochastic nature of gene expression \cite{Ozbudak02}, or due to 
protein conformational changes \cite{English06}. 

Hence, we model extrinsic random fluctuations 
on the parameters ${\boldmath{\mbox{$\mu$}}}$ by setting
for $i=1,...,R$:
\be
\mu_i \rightarrow \mu_i(t) = \bar{\mu}_i + \xi_i(t), \label{mustoch}
\ee
where $\bar{\mu}_i$ is the average value
of $\mu_i$, and the stochastic process $\xi_i(t)$ satisfies the equation
\be
\frac{d \xi_i}{dt} = -\frac{\xi_i}{\tau_i} +
\frac{\varepsilon_i^{1/2}}{\tau_i} \zeta_i(t). \label{ou} 
\ee
Eq. (\ref{ou}) defines the
well-known colored Ornstein-Uhlenbeck noise \cite{Gardiner97}. The
parameters $\tau_i$ play the role of the inverse correlation time of
the noise, while the parameters $\varepsilon_i$ are the noise intensities.
These account for different parameters to respond with
different sensitivities to the external fluctuations. The case when
one or more parameters do not fluctuate at all, is selected by
choosing the appropriate $\varepsilon_i$'s equal to
zero.  Furthermore, the assumption that 
\be
\varepsilon_i(\mu_i) \rightarrow 0 \quad {\rm as} \quad \mu_i
\rightarrow 0 \quad {\rm for} \quad i=1,...,R
\ee
guarantees that $\mu_i $ stays positive, and centered around $\bar{\mu}_i$.
Finally the noises $\zeta_i(t)$ 
are assumed to be Gaussian, with zero average, and correlations given
by
\be
\langle \zeta_i(t) \zeta_j(t^\prime) \rangle = 2 \delta_{ij} \delta(t
- t^\prime),  
\quad i,j=1,...,R. \label{noiseold}
\ee
By performing a linearization of the function
${\mathbf{f}}(\mathbf{c}, {\boldmath{\mbox{$\mu$}}})$ around
${\boldmath{\mbox{$\bar{\mu}$}}} = \{\bar{\mu}_1,...,\bar{\mu}_R\}$,
Eq. (\ref{determ}) is replaced by 
\be
\frac{d c_i}{dt} =
f_i(\mathbf{c},\boldmath{\mbox{$\bar{\mu}$}}) + 
\sum_{j=1}^R g_{ij}(\mathbf{c},{\boldmath{\mbox{$\bar{\mu}$}}})
\xi_j(t), \qquad i=1,...,N\label{sde0} 
\ee
with 
\be
g_{ij}(\mathbf{c},{\boldmath{\mbox{$\bar{\mu}$}}}) = \left. \frac{\partial
  f_i}{\partial \mu_j} \right|_{{\boldmath{\mbox{$\mu$}}} 
= {\boldmath{\mbox{$\bar{\mu}$}}}}.
\ee
Eq. (\ref{sde}) represents a stochastic differential equation driven by
multiplicative Ornstein-Uhlenbeck noise. For the sake of simplicity
we make the further approximation that the noise correlation times $\tau_i$'s
be much smaller than any other time scale in the system. In
this case the noises can be considered as virtually white. The process
originally defined by (\ref{determ}), (\ref{mustoch}), and
(\ref{ou}) becomes then to the first order in $\varepsilon_i$ and zero order in
$\tau_i$:
\be
\frac{d c_i}{dt} =
f_i(\mathbf{c},\boldmath{\mbox{$\bar{\mu}$}}) + 
\sum_{j=1}^R \varepsilon^{1/2}_j g_{ij}(\mathbf{c},{\boldmath{\mbox{$\bar{\mu}$}}})
\xi_j(t), \qquad i=1,...,N\label{sde} 
\ee
with 
\be
\langle \xi_i(t) \xi_j(t^\prime) \rangle = 2 \delta_{ij} \delta(t - t^\prime), 
\quad i,j=1,...,R. \label{noise}
\ee

Assuming the
noise as virtually white is legitimate in all those
cases when fluctuations are rapid, so that the zero $\tau_i$ expansion is
justified. This may be the case in many different experimental
situations, as it is highlighted for instance in \cite{Kar03}. Also
enzyme activities fluctuations as related to conformational changes have been
proven recently to have fast components \cite{English06,Ishikawa08}. 
Environmental fluctuations, such as light fluctuations, 
can be experimentally
realized so as to be very rapid as well. On the contrary, 
if fluctuations are not faster than the other
typical processes of the system, the present scheme breaks down.
In this case higher order terms in $\tau_i$ should be included to capture the full
dynamics, as it is in any approximation scheme. However, the
noise properties that we aim at discussing in this paper
emerge already at the zeroth order in $\tau_i$. Our scope is to design a minimal
model that accounts for the non-trivial properties of extrinsic noise.

As it is well known, the stochastic integral associated to the
multiplicative noise term in (\ref{sde}) is not uniquely defined when
white noise is assumed \cite{Gardiner97}, and its evaluation needs to
be supplemented with a prescription on how to define the discretized form of
(\ref{sde}). The assumption that the white noise (\ref{noise}) is the limit of
a correlated noise when the correlation time goes to zero resolves the
ambiguity, and leads naturally to the adoption of the so-called Stratonovich
prescription \cite{Gardiner97}. The same prescription is also valid
when an elimination of fast variables is implemented, as in the
reduction to the Michaelis-Menten form, as long as the
equilibration time of the fast variables is anyway smaller than the
correlation time of the noise \cite{Kupferman04}. 

A useful strategy to solve (\ref{sde}) is to derive the corresponding
Fokker-Planck equation \cite{vanKampen97}, and then compute the stationary probability
density function. The Stratonovich form of the Fokker-Planck equation 
for the stochastic process (\ref{sde}) with the correlation (\ref{noise}) is
\be
\frac{\partial p(\mathbf{c},t)}{\partial t} = - \sum_i
\frac{\partial}{\partial c_i}
\left\{\big[f_i(\mathbf{c},\boldmath{\mbox{$\bar{\mu}$}})
+ G_i(\mathbf{c},\boldmath{\mbox{$\bar{\mu}$}})\big]
p(\mathbf{c},t)\right\} + \sum_{i,j,k} \frac{\partial^2}{\partial c_i \partial
  c_j}\left\{\big[\varepsilon_k g_{ik}(\mathbf{c},\boldmath{\mbox{$\bar{\mu}$}})
  g_{jk}(\mathbf{c},\boldmath{\mbox{$\bar{\mu}$}})\big] 
  p(\mathbf{c},t)\right\},   
\ee
where $p(\mathbf{c},t)$ is a short notation for
$p(\mathbf{c},t|\mathbf{c_0},t_0)$, and 
\be
G_i(\mathbf{c},\boldmath{\mbox{$\bar{\mu}$}}) &=& \sum_{j=1}^R \sum_{k=1}^N
\varepsilon_j
\frac{\partial
  g_{ij}(\mathbf{c},\boldmath{\mbox{$\bar{\mu}$}})}{\partial c_k} 
g_{kj}(\mathbf{c},\boldmath{\mbox{$\bar{\mu}$}}) = \sum_{j=1}^R
\sum_{k=1}^N \varepsilon_j \left\{
\frac{\partial}{\partial c_k} \left[\frac{\partial}{\partial \mu_j}
  \sum_{l=1}^M S_{il} v_l \right] \right\} \left
\{\frac{\partial}{\partial \mu_j} 
  \sum_{h=1}^M S_{kh} v_h \right \} \label{stratdrift}
\ee
is known as the Stratonovich drift, here specified for the kinetics (\ref{kin}). 

Under the assumption of natural boundaries \cite{Gardiner97}, 
the Stratonovich stationary probability distribution can 
be evaluated to be:
\be
p_s(\mathbf{c}) = 
\exp \left\{ \int^\mathbf{c} \sum_i dc_i \sum_k
  D_{ik}^{-1}(\mathbf{c},\boldmath{\mbox{$\bar{\mu}$}})
  \left[\tilde{f}_k(\mathbf{c},\boldmath{\mbox{$\bar{\mu}$}}) - \sum_j  
  \frac{\partial}{\partial c_j}
  D_{kj}(\mathbf{c},\boldmath{\mbox{$\bar{\mu}$}}) \right] \right \},
\label{pst} 
\ee
where 
\be
D_{ij}(\mathbf{c},\boldmath{\mbox{$\bar{\mu}$}}) = \sum_k \varepsilon_k
g_{ik}(\mathbf{c},\boldmath{\mbox{$\bar{\mu}$}})
g_{jk}(\mathbf{c},\boldmath{\mbox{$\bar{\mu}$}}) \quad 
{\rm and} \quad
\tilde{f}_i(\mathbf{c},\boldmath{\mbox{$\bar{\mu}$}}) = 
f_i(\mathbf{c},\boldmath{\mbox{$\bar{\mu}$}}) + 
G_i(\mathbf{c},\boldmath{\mbox{$\bar{\mu}$}}).  
\ee

The maxima of the stationary probability
distribution $p_s(\mathbf{c})$ are assumed to correspond to the
macroscopic stable steady states of the pathway when noise is present,
as they represent 
the states where the pathway is found for most of the time. Other
indicators, such as the first moment, do not necessarily reflect
well the behaviour of the system \cite{Lefever84}. For this reason we shall focus on
calculating the most probable states of the pathway, under the
assumption that they represent the proper continuation of the
deterministic states when noise is switched on. 
By differentiating eq. (\ref{pst}) the extrema $\mathbf{c}_m$ of the probability
distribution are given by the expression:
\be
\mathbf{f}(\mathbf{c}_m, \boldmath{\mbox{$\bar{\mu}$}}) -
\mathbf{G}(\mathbf{c}_m,\boldmath{\mbox{$\bar{\mu}$}}) = 0. \label{condit}
\ee
The original fluctuations on parameters 
do not simply appear on the concentrations 
as mere fluctuations around their zero-noise 
values. Rather their effect is that of first shifting 
the value of the steady state concentrations, and then superimposing
fluctuations upon them. This mechanism originates directly from the
assumption of the noise being external. The small correlation
present (in any case smaller than the other typical time scales of
the system) makes the noise develop correlations with the
dynamical variables at the same time \cite{Carrillo03}. As a consequence the
multiplicative noise term in eq. (\ref{sde}) develops a non-zero
average. The Stratonovich prescription is a way of extracting the
non-zero average contribution in the limit of $\tau_i$ zero, and the 
corresponding modification of the 
dynamics. This modification is reflected in the change of the
probability distribution with respect to the deterministic case, as
well as of its zeroes, as shown by the condition (\ref{condit}). 
In general this accounts for a change in position, number, and
stability properties of the steady states with respect to the deterministic
solution.

\section{3. Implications for Metabolic Control Analysis} \label{impmca}

\subsection{3a. Standard Metabolic Control Analysis}

Metabolic Control Analysis (MCA) is a powerfull framework to relate 
local variables, such as enzymes activities or rate constants, to 
systemic variables, such as fluxes or concentrations. It focuses on 
establishing how the first exert control on the latter. 

Ideally, MCA is based on a simple sensitivity analysis, which 
can be schematized as follows: {\it i)} Let the system
relax to its steady state; {\it ii)} Apply a small perturbation; 
{\it iii)} Wait for relaxation onto the new steady state; {\it iv)} 
Measure the change in global variables (fluxes, concentrations, or
other systemic variables).  

The global effect of local changes is well described by a set of control
  coefficients, such as flux control coefficients,
\be
C_{v_j}^{J_i} = \frac{v_j}{J_i}\frac{\partial J_i}{\partial v_j} =
  \frac{\partial \ln J_i}{\partial \ln v_j}, \label{cj}
\ee
and concentration control coefficients, 
\be
C_{v_j}^{c_i} = \frac{v_j}{c_i}\frac{\partial c_i}{\partial v_j} =
  \frac{\partial \ln c_i}{\partial \ln v_j}. \label{cc}
\ee
These can be derived in the approximation of small perturbations from
the kinetic description of the metabolic network, 
for any given pathway, either branched or unbranched
\cite{Heinrich96}. Notice that the form of control coefficients
presented here, eqs. (\ref{cj}) and (\ref{cc}), contains the
normalization factors $v_j/J_i$ and $v_j/c_i$. This form allows the
characterizations of fractional changes and is suitable
for the cases when none of the $J_i$s or $c_i$s is zero, which might
happen in principle for certain combinations of the parameters. In
that case unnormalized control coefficients may be used \cite{Heinrich96}.

Control coefficients describe well the properties of the
system as a whole. 
%For instance in the case of concentration
%coefficients, it is intuitively clear that changing one particular 
%enzyme activity at a given step, produces a change in concentration of a
%particular metabolite. The new concentration however has
%to take into account the full kinetics
%and the rearrangement of all other concentrations as
%well. In this sense control coefficients describe systemic properties. 
This particular fact is expressed through the so-called Summation
Theorems, which relate coefficients for both concentrations and fluxes
to each other:
\be
\sum_{j=1}^M C_{v_j}^{c_i} = 0, \label{scc}
\ee
and 
\be
\sum_{j=1}^M C_{v_j}^{J_i} = 1. \label{sfc}
\ee
Equation (\ref{scc}) states that for any metabolite in the network 
the sum of the concentration control coefficients of all rates equals
zero. Similarly, for any flux the sum of all flux
control coefficients equals 1, as implied by eq. (\ref{sfc}). 
The fact that these sums are constrained demonstrates that
control is shared among all rates in the pathway, and highlights the
global properties of the network.
 
Local properties are equally important. In particular the response of
individual rates to the perturbation of both internal and external
variables may be very
informative. This sensitivity is expressed in terms of so-called
elasticity coefficients, whose definition is very similar to that of
control coefficients:
\be
\pi^{v_i}_{z} = \frac{z}{v_i}\frac{\partial
  v_i}{\partial z} = \frac{\partial
  \ln v_i}{\partial \ln z} . 
\ee 
Here $z$ can be a dynamical variable, such as an internal metabolic
concentration, or an external parameter, such as a reaction constant
or a fixed external concentration. When  $z$ is an internal metabolic
concentration, Connectivity Theorems for both fluxes 
and concentrations can be derived \cite{Klipp05}:
\be
&&\sum_{l=1}^M C_{v_l}^{J_i} \pi^{v_l}_{c_j} = 0,\\
&&\sum_{l=1}^M C_{v_l}^{c_i} \pi^{v_l}_{c_j} = - \delta_{ij}.
\ee 
Here $\delta_{ij}$ is the Kronecker symbol, equal to 1 for $i=j$, and
0 otherwise. Together with Summation Theorems, Connectivity Theorems
allow for the calculation of the control coefficients from the
knowledge of the elasticities, and therefore integrate between a local
and a global description of the network.

In the present context we are also interested in the case when $z$ is
a parameter, such as any of the $\mu_j$s introduced in the previous
section. Perturbation of parameters is described at the systemic
level in terms of response coefficients, such as concentration
response coefficients,
\be
R^{c_i}_{\mu_j} = \frac{\mu_j}{c_i}\frac{\partial
  c_i}{\partial \mu_j} = \frac{\partial
  \ln c_i}{\partial \ln \mu_j}, 
\ee  
and flux response coefficients,
\be
R^{J_i}_{\mu_j} = \frac{\mu_j}{J_i}\frac{\partial
  J_i}{\partial \mu_j} = \frac{\partial
  \ln J_i}{\partial \ln \mu_j}. 
\ee  
The interplay among response coefficients, control coefficients, and
elasticities is formalized by the Partitioned Response relations, for both
concentrations and fluxes, 
\be
R^{c_i}_{\mu_j} = \sum_{i=1}^M C^{c_i}_{v_i} \pi^{v_i}_{\mu_j} \quad\quad
{\rm and} \quad\quad R^{J_i}_{\mu_j} = \sum_{i=1}^M C^{J_i}_{v_i} \pi^{v_i}_{\mu_j}.
\label{pr}
\ee
Eqs. (\ref{pr}) show that the system response
originates locally through the sensitivity of local rates to the
external perturbation, and is transfered to the systemic level 
through the mediation of the control that each rate
has on the system variable of interest. Response coefficients are therefore
determined by both local sensitivities to external effectors, and
control exerted by individual rates. The set of control  
coefficients, response coefficients and elasticities, obeying Summation
and Connectivity Theorems and/or Partitioned Response, represents a useful
framework for connecting local and global properties of metabolic
networks. In the next section we shall see that this mathematical structure is
robust to extrinsic noise, but at the same time is sensitive to it through the 
acquired explicit dependency of all coefficients on the noise intensity.

\subsection{3b. Metabolic Control Analysis with
  extrinsic stochasticity} \label{summcon}
One way to derive control, response, and elasticity 
coefficients, and their structural relationships, 
is to use the perturbative procedure mentioned
in the previous section, and work out the corresponding linear
approximation for small perturbations \cite{Hofmeyr01}. 
However, there is another elegant and compact way of
deriving such properties when concentrations and rates are homogeneous
functions.

Say that $f$ is a function of $n$ variables $x_1,...x_n$, 
which satisfies a scaling relationship, such as:
\be
f(\lambda^{\alpha_1}x_1,...,\lambda^{\alpha_n}x_n) = \lambda^\gamma
f(x_1,...,x_n). 
\ee
Then it is possible to prove that the following equality holds true:
\be
\sum_{i=1}^n \alpha_i \frac{\partial \ln f}{\partial \ln x_i} =
\gamma, \label{rel} 
\ee 
and $f$ is said to be a homogeneous function of degree $\gamma$ (Euler's Theorem).
Homogeneity of concentrations and rates, of degree 0
and 1 respectively, has been shown to lead to
Summation and Connectivity Theorems \cite{Giersch88a,Giersch88b}. 

In order to follow this approach, let us consider the equation that
determines the maximum probability concentrations, eq. (\ref{condit}): 
\be
\sum_{l=1}^M S_{il} v_l - 
\sum_{j=1}^R \sum_{k=1}^N \sum_{l=1}^M \sum_{h=1}^M 
\varepsilon_j 
\left[ \frac{\partial}{\partial c_k} \left(\frac{\partial v_l}{\partial \mu_j}
  \right) \right] \left[
\frac{\partial v_h}{\partial \mu_j} \right] 
   S_{il} S_{kh} = 0, \quad i=1,...,N. \label{steady0}
\ee
It is not restrictive to assume specificity of the parameters 
with respect to the rates, namely 
\be
\frac{\partial v_l}{\partial \mu_j} \neq 0 \quad \Rightarrow \quad
\frac{\partial v_h}{\partial \mu_j} = 0 \qquad \forall h \neq l.
\label{spec} 
\ee
This allows us to get rid of the sum over $h$ in
(\ref{steady0}). Also, by 
defining $R_l$ as the number of parameters entering in the
kinetics of $v_l$, 
with $R=R_1 + R_2 + ... + R_M$, eq. (\ref{steady0}) can be rewritten
in the more compact form
%\be
%\sum_{l=1}^M S_{il} v_l - 
%\sum_{l=1}^M \sum_{j=1}^{R_l} \sum_{k=1}^N \varepsilon_j 
%\left[ \frac{\partial}{\partial c_k} \left(\frac{\partial v_l}{\partial \mu_j}
%  \right) \right] \left[
%\frac{\partial v_l}{\partial \mu_j} \right] 
%   S_{il} S_{kl} = 0, \qquad i=1,...,N. \label{steady}
%\ee
\be
\sum_{l=1}^M S_{il} \tilde{v}_l = 0, \qquad i=1,...,N, 
\label{compi2} 
\ee
where
\be
\tilde{v}_l = v_l - v_{l}^\varepsilon, \qquad
l=1,...,M \label{nr}
\ee
and
\be
v_{l}^\varepsilon = \sum_{j=1}^{R_l} \sum_{k=1}^N
\varepsilon_j 
\left[ \frac{\partial}{\partial c_k} \left(\frac{\partial v_l}{\partial \mu_j}
  \right) \right] \left[
\frac{\partial v_l}{\partial \mu_j} \right] 
   S_{kl}, \qquad l=1,...,M.  \label{vlj}
\ee

Equation (\ref{compi2}) allows for a particularly transparent
interpretation. Every rate present in the
deterministic system $v_l$ appears together with the Stratonovich drift
$v_{l}^\varepsilon$, made up of as many terms as the number $R_l$ of parameters
affected by noise that enter the definition of $v_l$. 
Notice that in the case when the dependency on concentrations and parameters
affected by noise is linear, the drifts (\ref{vlj}) have the same
analytical form as the zero noise rates $v_l$.
In this case the noise simply causes a renormalization of 
parameter values. In contrast, for general nonlinear dependencies,
eq. (\ref{vlj}) defines interaction terms not necessarily of
the same form as those defining the zero noise system. 

Requiring that concentrations be homogeneous functions of degree
0 is equivalent to require the invariance of (\ref{compi2}) under
rescaling of the rates. In turn this rescaling needs to be 
defined consistently so that the rates themselves be homogeneoeus of degree 1.
In this way, while individual rates undergo a scale tansformation,
concentrations will stay invariant. 

It is clear that the scaling properties of the 
Stratonovich drift, eq. (\ref{vlj}), depend on the specific form of
the deterministic rates, and on the
parameter that has been perturbed by noise. We shall focus on three
different cases, as representative of most situations in metabolic
networks: {\em i)} rates obeying the law of mass action, {\em ii)}
rates described by Michaelis-Menten kinetics, and {\em iii)}
rates described by reversible Michaelis-Menten kinetics.

Consider first a rate, $v_{\rm lin}$, having the form given by the law
of mass action, and let $\mu_{\rm lin}$ be one of its parameters
affected by noise. Since $v_{\rm lin}$ is linear in $\mu_{\rm lin}$, 
the rescaling $\mu_{\rm lin} \rightarrow \lambda \mu_{\rm lin}$
implies $v_{\rm lin} \rightarrow \lambda v_{\rm lin}$. 
Then the associated Stratonovich drift $v_{\rm lin}^\varepsilon $ 
can be made to scale as $v_{\rm lin}$ only if $\varepsilon_{\mu}
\rightarrow \lambda \varepsilon_{\mu}$ 
where $\varepsilon_{\mu}$ is the intensity of the noise acting 
upon $\mu_{\rm lin}$. Therefore the rate $\tilde{v}_{\rm lin}$ scales as
$\lambda$, that is it is homogeneous of degree 1, if 
\be
v_{\rm lin} \rightarrow \lambda v_{\rm lin} \quad \mbox{and} \quad
\varepsilon_{\mu} \rightarrow \lambda \varepsilon_{\mu}. 
\label{scaling}
\ee

Next consider rates of the irreversible
Michaelis-Menten form, $v_{\rm MM} = V_{\rm max} S/(K_m + S)$,
with $S$ the substrate concentration, and $V_{max}$ and $K_m$
maximal velocity and Michaelis-Menten constant respectively. The
general form of the scaling properties of $v_{\rm MM}$ is
\be
v_{\rm MM}(\lambda^\alpha V_{\rm max}, \lambda^0 K_m) =  \lambda^\alpha
v_{\rm MM}(V_{\rm max}, K_m).  
\ee
This implies $v_{\rm MM} \rightarrow \lambda v_{\rm MM}$, when $V_{\rm max}
\rightarrow \lambda V_{\rm max}$ 
and $ K_m \rightarrow K_m$. To find the scale transformation of the
associated Stratonovich drift $v_{\rm MM}^{\varepsilon}$, we consider
separately the two cases, 
when the noise acts either upon $V_{\rm max}$, or upon $K_m$. If
$V_{\rm max}$ is affected by a noise having intensity $\varepsilon_{V}$, 
the corresponding $v_{\rm MM}^\varepsilon $ scales as $v_{\rm MM}$
again if $\varepsilon_{V}$ scales as $\lambda$. So that the rate
$\tilde{v}_{\rm MM}$ scales as $\lambda$ if 
\be
v_{\rm MM} \rightarrow \lambda v_{\rm MM} \quad \mbox{and} \quad
\varepsilon_{V} \rightarrow \lambda \varepsilon_{V}. \label{scalingbis}
\ee
Notice that (\ref{scalingbis}) is the same set of scale transformations as
(\ref{scaling}). This is not surprising since the dependency of any
Michaelis-Menten type of rate on maximal velocities is in fact linear,
and therefore the previous case must apply. 

In contrast, if $K_m$ is the parameter of $v_{\rm MM}$ perturbed by a noise with
intensity $\varepsilon_{K}$, 
then the corresponding $v_{\rm MM}^\varepsilon $ scales as
$\lambda^2 \varepsilon_{K}$ when   
$V_{\rm max} \rightarrow \lambda V_{\rm max}$ and $ K_m \rightarrow
K_m$. Therefore, imposing $\varepsilon_{K}$ to scale as
$\lambda^{-1}$ makes $v_{\rm MM}^\varepsilon $ 
to scale as $\lambda$, which is what we want in order for
$\tilde{v}_{\rm MM}$ to be homogeneous of degree 1:
\be
v_{\rm MM} \rightarrow \lambda v_{\rm MM} \quad \mbox{and} \quad
\varepsilon_{K} \rightarrow \lambda^{-1} \varepsilon_{K}.
\ee 

The same result can be obtained also in the case of the
  reversible Michaelis-Menten rate,
\be
v_{\rm RMM} = \frac{\frac{V_{\rm max}^{\rm f}}{K_S} S -
  \frac{V_{\rm max}^{\rm b}}{K_P} P}{1 + \frac{S}{K_S} + \frac{P}{K_P}}, 
\ee
where $V_{\rm max}^{\rm f}$ and $V_{\rm max}^{\rm b}$ are maximal
  velocities for the forward and backward reactions respectively, and
  $K_S$ and $K_P$ are the Michaelis-Menten constants associated to the
  substrate $S$ and the product $P$. 
The scaling properties of $v_{\rm RMM}$ are similar to the
  irreversible case:
\be
v_{\rm RMM}(\lambda^\alpha V_{\rm max}^{\rm f},
  \lambda^\alpha V_{\rm max}^{\rm b}, \lambda^0 K_S, \lambda^0 K_P ) =  \lambda^\alpha
v_{\rm RMM}(V_{\rm max}^{\rm f},V_{\rm max}^{\rm b}, K_S, K_P).  
\ee
As before, if $V_{\rm max}^{\rm f}$ and/or $V_{\rm max}^{\rm b}$
  are affected by noise, $\tilde{v}_{\rm RMM}$ scales as $\lambda$ if  
\be
v_{\rm RMM} \rightarrow \lambda v_{\rm RMM} \quad \mbox{and} \quad
\varepsilon_{V} \rightarrow \lambda \varepsilon_{V}. 
\ee
On the other hand, if any of the Michaelis-Menten constants are
  affected by noise, then the scaling   
\be
v_{\rm RMM} \rightarrow \lambda v_{\rm RMM} \quad \mbox{and} \quad
\varepsilon_{K} \rightarrow \lambda^{-1} \varepsilon_{K}.
\ee 
makes sure that $\tilde{v}_{\rm RMM}$ is homogeneous of degree 1.

In summary, under rescaling $v_l \rightarrow \lambda v_l$, 
the rates $\tilde{v}_l$ are homogeneous functions of
degree 1 if the noise intensities of any parameter appearing
linearly in the rates are rescaled as $\lambda$ (including
maximal velocities of possible Michaelis-Menten
rates), and the noise intensities associated to Michaelis-Menten
constants are rescaled as $\lambda^{-1}$. Also, because of the
invariance of (\ref{compi2}) under this same rescaling, the maximum 
probability concentrations are homogeneous functions of degree $0$.
For other possible types of parameter dependencies, the proper scale
transformation need to be sought case by case by applying similar arguments.

By using Euler's Theorem, the homogeneity of concentrations and rates
leads directly to the Summation Theorems for the respective control
coefficients. In particular, from the scaling relation
\be
{\boldmath{\mbox{$c$}}}^m(\lambda \tilde{v}_1,..., \lambda \tilde{v}_M) =
{\boldmath{\mbox{$c$}}}^m(\tilde{v}_1,...,\tilde{v}_M),
\ee
and by using (\ref{rel}), we readily obtain
\be
\sum_{j=1}^M \frac{\partial \ln c_i^m}{\partial \ln \tilde{v}_j} = 0,
\label{sumcon} 
\ee
that is 
\be
\sum_{j=1}^M C_{\tilde{v}_j}^{c_i^m} = 0. \label{sumconcomp}
\ee
Eq. (\ref{sumconcomp}) proves that the Summation Theorem 
for concentration control coefficients is still valid when stochasticity is
present in the form of noise on parameters.

Similarly, by defining the macrosopic fluxes as the fluxes evaluated in 
correspondence of the most probable
concentrations,
${\boldmath{\mbox{$J$}}}^m =
\tilde{{\boldmath{\mbox{$v$}}}}(\mathbf{c}^m)$, 
it is clear that
\be
{\boldmath{\mbox{$J$}}}^m(\lambda
\tilde{v}_1,...,\lambda \tilde{v}_M) = \lambda {\boldmath{\mbox{$J$}}}^m
(\tilde{v}_1,...,\tilde{v}_M).  
\ee
Euler's Theorem may be invoked again, and implies
\be
\sum_{j=1}^M C_{\tilde{v}_j}^{J_i^m} = 1, \label{stjc}
\ee
which states that the Summation Theorem for flux control coefficients
is valid as well, even in presence of extrinsic noise. 

The substitution of the zero noise rates $v_l$ with the rates
(\ref{nr}), which obey the same scaling
properties, allows also to derive in a standard fashion Connectivity Theorems
\cite{Giersch88b} and Partitioned Response \cite{Hofmeyr01}: 
\be
\sum_{l=1}^M C_{\tilde{v}_l}^{J_i^m} \pi^{\tilde{v}_l}_{c_j^m} = 0
\qquad {\rm and} \qquad  
\sum_{l=1}^M C_{\tilde{v}_l}^{c_i^m} \pi^{\tilde{v}_l}_{c_j^m} = - \delta_{ij}.
\label{cr}
\ee 
and
\be
R^{c_i}_{\mu_j} = \sum_{i=1}^M C^{c_i^m}_{\tilde{v}_i}
\pi^{\tilde{v}_i}_{\mu_j} \quad\quad 
{\rm and} \quad\quad R^{J_i^m}_{\mu_j} = \sum_{i=1}^M
C^{J_i}_{\tilde{v}_i} \pi^{\tilde{v}_i}_{\mu_j}. 
\label{prr}
\ee
It is interesting to note that new elasticity coefficients can be defined, which
describes the local effect of perturbations of the noise intensities,
\be
\pi^{\tilde{v}_i}_{\varepsilon_j} 
= \frac{\varepsilon_j}{\tilde{v}_i}
\frac{\partial \tilde{v}_i}{\partial \varepsilon_j},
\ee
together with new response coefficients, for both concentrations and
fluxes, which in turn describe the effect of the noise on the
sytemic variables: 
\be
R^{c_i^m}_{\varepsilon_j} = \frac{\varepsilon_j}{c_i^m}\frac{\partial
  c_i^m}{\partial \varepsilon_j}
\qquad {\rm and} \qquad R^{J_i^m}_{\varepsilon_j} =
\frac{\varepsilon_j}{J_i^m}\frac{\partial 
  J_i^m}{\partial \varepsilon_j}.
\ee  

Summation Theorems (\ref{sumconcomp}) and (\ref{stjc}), 
Connectivity Theorems (\ref{cr}), and Partitioned Response (\ref{prr})
show that the mathematical structure of the network 
is robust against extrinsic noise. This property is non 
trivial, because it might be expected that
extrinsic noise might break the homogeneity of the
system. Partitioned Response originates in fact from the breakdown 
of homogeneity of fluxes, but in the same way as in the deterministic
case. The fact that the Stratonovich drift obeys the same scale
transformation as the zero noise rates guarantees that these properties
are preserved. However, it should be noted that individual control
coefficients, as well as response coefficients and elasticities, 
depend explicitly on the noise intensity. This feature
demonstrates that extrinsic noise can in fact act as a control
mechanism. It modifies the local rates, as formalized
by the corresponding elasticy coefficients, and propagates up to the
systemic level to affect global variables.

\section{4. An explicit example}\label{example}

\begin{figure}[t]
\begin{center}
\resizebox{0.2\columnwidth}{!}{
\includegraphics{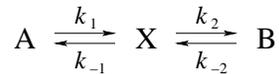}}
\caption{The elementary reaction pathway chosen as an
  illustration of the effect of extrinsic fluctuations.} 
\label{elem}
\end{center}
\end{figure}

As an example of the theory presented, let us consider the 
chemical reaction depicted in Fig. \ref{elem}.
Despite its simplicity it serves well the purpose of illustrating the
notion of control by noise.

Let us define the concentrations of the species $A$, $B$, and $X$ as 
$a$, $b$, and $x$ respectively. Let us assume that the
constant $k_2$ is the only parameter undergoing stochastic fluctuations.
This leads to the following Stratonovich SDE 
\be
\frac{dx}{dt} = v_1 - v_2 - \varepsilon^{1/2} x \xi(t), \label{exSt}
\ee
where $\xi(t)$ is Gaussian white noise. The deterministic reactions
$v_1$ and $v_2$ are given in
terms of the rate constants defined in Fig. \ref{elem}{\em a} as 
\be
&&v_1 = k_1 a  - k_{-1} x, \label{nu1}\\ 
&&v_2 = k_2 x - k_{-2} b.\label{nu2} 
\ee

Notice that Equation (\ref{exSt}) is exact at
all orders in $\varepsilon$, because of the linear dependence of the
kinetics on $k_2$. In general, this dependence may be nonlinear, and the
equation corresponding to (\ref{exSt}) in the nonlinear case should be considered 
valid only at the leading order in $\varepsilon$. 

By following Section 2, we compute the
stationary concentrations and fluxes in order to derive 
the corresponding control coefficients. In one dimension, with
extrinsic fluctuations acting upon only one parameter, 
eq. (\ref{pst}) becomes:
\be
p_s(x) = \frac{\cal N}{g(x)} \exp \left( \frac{1}{\varepsilon} \int^x
\frac{f(x^\prime)}{g^2(x^\prime)} dx^\prime \right), 
\ee
with $f(x) = v_1(x) - v_2(x)$, $g(x) = -x$ and ${\cal N}$ a
normalization constant that can be directly calculated by imposing
$\int_o^\infty p_s(x) dx = 1$. The explicit expression of $p_s(x)$ is then
\be
p_s(x) = \left(\frac{k_1 a + k_{-2} b}
{\varepsilon}\right)^{(k_{-1}+k_2)/
\varepsilon} \frac{x^{- \left(1 + (k_{-1} +
    k_2)/\varepsilon\right)}}{\Gamma((k_{-1} + k_2)/\varepsilon)} 
\exp\left(-\frac{k_1 a + k_{-2} b}{\varepsilon x}\right). \label{pdf}
\ee
The extrema of $p_s(x)$, as from eq. (\ref{condit}), result in 
$v_1 - v_2 - \varepsilon x = 0$, which gives
\be
x_m = \frac{k_1 a + k_{-2} b}{k_{-1} + k_2 + \varepsilon}. \label{staz}
\ee
Notice that all the moments of the distribution (\ref{pdf}) can be
calculated, and in particular we have:
\be
\langle x \rangle = \int_0^\infty dx \;x \; p_s(x) = \frac{k_1 a + k_{-2}
  b} {k_{-1} + k_2 - \varepsilon}, \qquad \varepsilon < k_{-1} + k_2 .
\label{aver}
\ee

\begin{figure}[t]
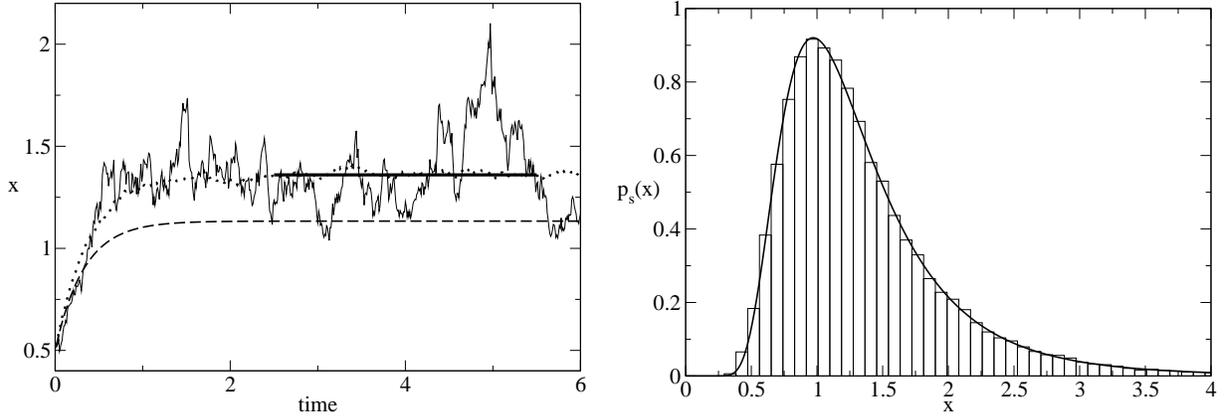

\begin{center}
\resizebox{0.9\columnwidth}{!}{
\includegraphics{conc.eps} \hspace{1cm}
\includegraphics{pdf.eps}}
\caption{{\em Left panel}: Dependency over time of the concentration $x$, as given by
  eq. (\ref{exSt}). Parameters have been set arbitrarily to the
  following values: $k_1=1\;{\rm min}^{-1}$, $k_{-1}=1\;{\rm min}^{-1}$, $k_{2}=2\;
  {\rm min}^{-1}$, $k_{-2}=1.2\;{\rm min}^{-1}$. External
  concentrations have been taken as $a=1$ mM and $b=2$ mM, while the
  noise intensity has been fixed to $\varepsilon = 0.5$. The dashed
  curve corresponds to the solution of the deterministic equation, the spiky and
  dotted lines show the value of the concentration averaged
  respectively over 10 and 1000 realizations of the noise. The thick
  straight line is the value of the average concentration has from
  eq. (\ref{aver}), namely $\langle x \rangle = 1.36$ mM with the
  chosen parameters. 
  {\em Right panel}: The stationary probability distribution (\ref{pdf})
  (solid line), versus the estimate coming from the numerical solution
  of eq. (\ref{exSt}), with the parameters as above. The maximum is
  at $x_m = 0.97$ mM, whereas the deterministic value is $x_m = 1.13$ mM. 
}\label{conc}
\end{center}
\end{figure}

Fig. \ref{conc} shows a direct numerical simulation of the SDE
(\ref{exSt}) with the Stratonovich prescription. The numerical
integration scheme is standard (see for instance
\cite{GarciaOjalvo99}). In the left panel of Fig. \ref{conc}
comparison is shown between the deterministic solution, namely the
solution of (\ref{exSt}) with the noise term switched off, the
full stochastic solution, and the expectation $\langle x \rangle$, 
averaged over 10 and 1000 realizations of the noise. The theoretical
value of $\langle x \rangle$ as given by (\ref{aver}) is also
plotted for comparison. In the right panel of the same figure we show
the probability density (\ref{pdf}) and its numerical evaluation from
the solution of eq. (\ref{exSt}). Notice that the maximum of the
probability density has undergone indeed a shift with respect to the
deterministic value, by changing its value from $x_m = 1.13$ mM
(deterministic) to $x_m = 0.97$ mM (stochastic). The same is true for the 
expectation $\langle x \rangle$, whose value has changed from $\langle x \rangle =
1.13$ mM (deterministic) to $\langle x \rangle = 1.36$ mM (stochastic). 

In summary, the effect caused by extrinsic fluctuations affecting 
the parameter $k_2$ is 
a shift in the value of the maximum of the probability density, which
induces a shift also on the value of its first moment. 
Notice that this mechanism corresponds effectively to 
resetting the value of the parameter $k_2$ to $k_2 +
\varepsilon$. This resetting is non-trivial in that it rests on 
the action of the fluctuations only, which shift the value of $k_2$ 
effectively perceived by the system. Nonetheless the bare $k_2$ is blocked 
to its zero-noise value. Due to the
linearity of the system, this is the only effect, and more subtle
modifications, such as in the number of stationary states of the 
probability distribution, or in their stability properties, are not
present. 

The same resetting appears in the expression for the flux as well,
which becomes thereby explicitly dependent on the noise intensity. In fact, 
the rates (\ref{nr}) read in this case:
\be
&&\tilde{v}_1 = k_1 a  - k_{-1} x,\\ 
&&\tilde{v}_2 = (k_2 + \varepsilon) x - k_{-2} b,
\ee
and their direct evaluation at the value $x_m$ results in: 
\be
&&J^m = J_1^m = J_2^m = \tilde{v}_1(x_m) = \tilde{v}_2(x_m) =
\frac{k_1 (k_2 + \varepsilon) a - k_{-1} 
  k_{-2} b}{k_{-1} + k_2 + \varepsilon}. \label{j}
\ee

The noise dependency of the maximum probability
concentration (\ref{staz}) and of the corresponding flux (\ref{j}) can
be interpreted within Metabolic Control Analysis. To this aim, we compute now
control coefficients, response coefficients and elasticies, and verify that the
Summation and Connectivity Theorems are fulfilled, as from Section 3b. 

The concentration control coefficients result in 
\be
&&C_{\tilde{v}_1}^x = \frac{k_1 (k_2 + \varepsilon) a - k_{-1} k_{-2}
  b}{(k_{-1} + k_2 + 
  \varepsilon)(k_1 a + k_{-2} b )}, \label{cxnu1}\\
&&C_{\tilde{v}_2}^x = - \frac{k_1 (k_2 + \varepsilon) a - k_{-1} k_{-2}
  b}{(k_{-1} + k_2 + 
  \varepsilon)(k_1 a + k_{-2} b )},\label{cxnu2}
\ee
and similarly for the flux $J^m$,
\be
&&C_{\tilde{v}_1}^{J} = \frac{k_2 +
  \varepsilon}{k_{-1} + k_2 + \varepsilon}, \label{cj1nu1}\\ 
&&C_{\tilde{v}_2}^{J} = \frac{k_{-1}}{k_{-1} + k_2 + \varepsilon}. \label{cj1nu2}
\ee
It is immediate to verify that the Summation Theorems hold true, as expected:
\be
C_{v_1}^x + C_{v_2}^x = 0  \qquad {\rm and} \qquad C_{v_1}^J + C_{v_2}^J = 1.
\ee

\begin{figure}[t]
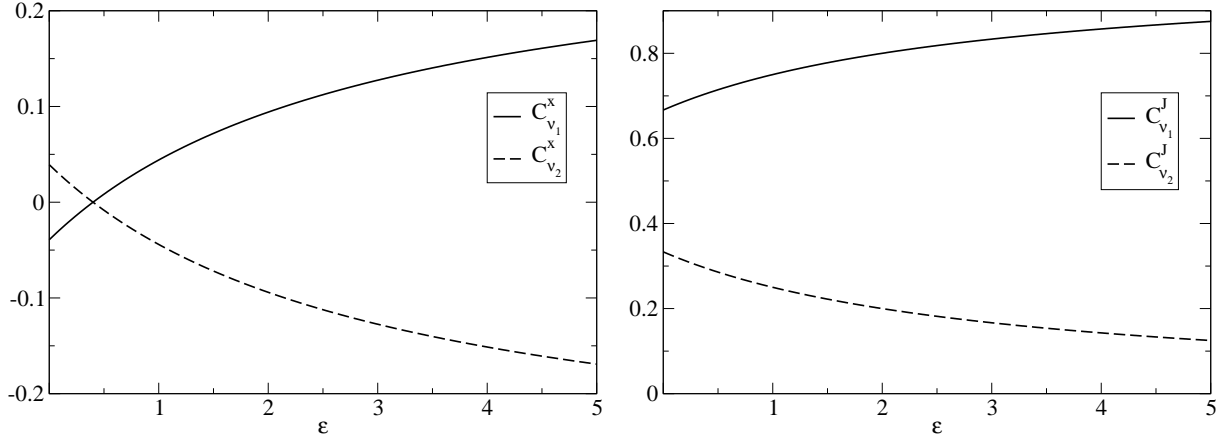

\begin{center}
\resizebox{0.9\columnwidth}{!}{
\includegraphics{CS.eps}\hspace{1cm}\includegraphics{CJ.eps}}
\caption{Concentration ({\em left panel}) and flux ({\em right
  panel}) control coefficients as from
  eqs. (\ref{cxnu1})-(\ref{cj1nu2}) as functions of
  $\varepsilon$. Parameters setting is as in
  the caption of Fig. \ref{conc}. 
}
\label{CJ}
\end{center}
\end{figure}

In Fig. \ref{CJ} we present a numerical
evaluation of expressions (\ref{cxnu1})-(\ref{cj1nu2}) as functions of
the noise intensity 
$\varepsilon$, with parameters chosen as indicated in the caption  
of Fig. \ref{conc}. As eqs. (\ref{staz}) and (\ref{j}) show, 
the effect of the noise is to allow an extra
tuning mechanism for modulating the systemic variables. Futhermore, noise
may also allow for a change of the control that 
individual rates exert on the whole pathway. In
other words, the dependency of the control coefficients on the noise 
makes it possible for the system to rebalance controls, and possibly
to explore dynamical regions that might not be accessible given the
standard parameter values. Already in this simple linear system, this effect
can be so dramatic as to invert the sign of the control of the two rates
on the concentration, as the left panel of Fig. \ref{CJ} shows. Of
course this extra tuning is always constrained by the respective
Summation Theorems.
 
As proven in section 3b, Connectivity Theorems are also preserved. It
is a simple matter to compute the elasticities,
\be
&&\pi_x^{\tilde{v}_1} = - \frac{k_1 k_{-1}a + k_{-1} k_{-2}b}{k_1(k_2 +
  \varepsilon) a - k_{-1} k_{-2} b},\\
&&\pi_x^{\tilde{v}_2} = \frac{k_1 (k_2 + \varepsilon) a + k_{-2}
  (k_2 + \varepsilon) b}{k_1(k_2 +
  \varepsilon) a - k_{-1} k_{-2} b},\\
\ee
and verify that
\be
C^x_{\tilde{v}_1} \pi_x^{\tilde{v}_1} + C^x_{\tilde{v}_2}
\pi_x^{\tilde{v}_2} = -1 \qquad {\rm and} \qquad
C^J_{\tilde{v}_1} \pi_x^{\tilde{v}_1} + C^J_{\tilde{v}_2} 
\pi_x^{\tilde{v}_2} = 0.
\ee
On the other hand, elasticities and response coefficients with respect to noise 
turn out to be 
\be
&&\pi_{\varepsilon}^{\tilde{v}_1} = 0,\\
&&\pi_{\varepsilon}^{\tilde{v}_2} = \frac{\varepsilon (k_1 a + k_{-2}
  b)}{k_1(k_2 + \varepsilon) a - k_{-1} k_{-2} b}, \label{elastp}  
\ee
and
\be
&&R^x_{\varepsilon} = - \frac{\varepsilon}{k_{-1} + k_2 + \varepsilon},\label{rp1}\\
&&R^J_{\varepsilon} = \frac{\varepsilon k_{-1}(k_1 a + k_{-2}
  b)}{(k_{-1} + k_2 + \varepsilon)(k_1(k_2 + \varepsilon) a - k_{-1}
  k_{-2} b)}, \label{rp2}
\ee
and obey the Partitioned Response:
\be 
R^x_{\varepsilon} = C^x_{\tilde{v}_1} \pi_{\varepsilon}^{\tilde{v}_1}
+ C^x_{\tilde{v}_2} \pi_{\varepsilon}^{\tilde{v}_2} = C^x_{\tilde{v}_2}
\pi_{\varepsilon}^{\tilde{v}_2} \qquad {\rm and} \qquad
R^J_{\varepsilon} = C^J_{\tilde{v}_1} \pi_{\varepsilon}^{\tilde{v}_1}
+ C^J_{\tilde{v}_2} \pi_{\varepsilon}^{\tilde{v}_2} = C^J_{\tilde{v}_2}
\pi_{\varepsilon}^{\tilde{v}_2}.
\ee
The elasticity (\ref{elastp}) highlights the local effect of noise,
and the response coefficients (\ref{rp1}) and (\ref{rp2}) for concentrations 
and fluxes are a manifestation of its global effect at the systemic
level. While the action of noise on control coefficients is only a modification of
the existiting ones, which become noise dependent, the elasticity
(\ref{elastp}) and the response coefficients (\ref{rp1}) and
(\ref{rp2}) are truly ``new'' coefficients, which are turned on by the presence of
noise. Nonetheless they obey Partitioned Response in the standard
fashion. In this sense, the mathematical structure of the theory is
generally preserved.

\section{5. Conclusion and outlook}\label{concl}

Metabolic Control Analysis is a fundamental framework 
that links both conceptually and practically local and global properties of metabolic
networks. In this paper we propose a
way of generalizing MCA, and its predictive power at the systemic level, 
when fluctuations are present in control parameters, and the
kinetics are affected by so-called extrinsic noise. 

The proposed generalization relies on the acquired explicit dependency
of all standard MCA coefficients on noise intensity, and on the definition of new
elasticity and response coefficients accounting for noise perturbations. 
Even though Summation and Connectivity Theorems, as well as
Partitioned Response, are shown to be robust against noise, noise
introduces an extra tuning mechanism, capable of modifying the local and
global properties of the pathway.

In fact, through the noise dependency of the MCA coefficients, we aim
at putting forward a new way of looking at noise in metabolic networks.  
We propose to go beyond the interpretation of noise
as a mere experimental nuisance, and to explore its potentiality
to exert external functional control onto the system. As a result of our
analysis, the behaviour of the pathway can be regulated
by the insertion of {\em ad hoc} random fluctuations on properly chosen
parameters. 

Intrinsic noise, related for instance 
to the low copy number of molecules or to protein conformational
fluctuations, can also be described by this approach.
From within a modelling perspective, deciding whether noise is
internal or external to the system is so delicate an issue as defining the
pathway of interest itself. In an integrated
description of the full network all noise has an internal origin. However, the description of
reduced modules, as opposed to the fully integrated approach,
requires the adoption of effective kinetics, where 
intrinsic noise is translated into extrinsic noise
on parameters. The predictions of these effective kinetics are the 
ones, which should be compared with experimental data, or to full
stochastic simulations.  

This is particularly relevant when the system is characterized 
by a time-scale separation between
slow and fast variables. If the latter are eliminated by standard
techiques \cite{Elf03}, their dynamical contribution is rewritten in
parametric form. Accordingly original intrinsic fluctuations acting on
fast variables are perceived as extrinsic noise in the reduced system.  
As long as the reduction procedure is carried out before the
approximation of white noise \cite{Kupferman04}, the reduced dynamics
is modified by the Stratonovich drift. This scenario is confirmed 
for instance in \cite{Elf03}, where an improved version of the Linear
Noise Approximation combined with the elimination of fast variables
in a simple metabolic network shows a very good agreement with the
results of a direct stochastic simulation \cite{Elf03}.

In all other cases intrinsic noise can be described by standard tools
such as Master Equations, and possibly their Fokker-Planck
\cite{vanKampen97} or Linear Noise approximations \cite{Elf03}. 
Effective Langevin descriptions can then be derived, but, apart from some
exceptions \cite{Carrillo03}, these
are generally characterized by zero average multiplicative noise
terms, with no Stratonovich drift associated. This implies that no
control mechanism can be expected. 

A recent work  by Levine and Hwa \cite{Levine07} confirms this prediction.
The Authors of \cite{Levine07} study 
the effect that intrinsic noise, as associated to copy number
fluctuations of metabolic species, may have on metabolic pathways. The
conclusion of the Authors is that noise ``does not propagate'',
being the fluctuations of the metabolites in the pathway uncorrelated from each
other. This is fully consistent with our framework. However possible correlations 
among metabolic fluctuations might in fact arise when fluctuations 
of enzymes common to two or more metabolites are considered. 
Our treatment of extrinsic noise aims to address this particular
case.

In this respect, the possible internal origin of extrinsic noise
suggests an even more interesting role of stochastic fluctuations. 
Robustness of biochemical systems to perturbations is often thought to
be a fundamental property, and selection of robust traits has
been proposed as a driving principle of evolution \cite{Kitano04}. 
Noise of course is no exception, and the capability of preserving 
functionality despite noise is certainly a fundamental feature of living
systems (see for instance \cite{Eldar02} for an example). 
A different perspective, however, emerges from considering 
cellular noise as a genetic response to chemical or physical constraints, 
as represented for instance by the values of reaction constants in a
biochemical network. As we have shown, the operational point of the network 
is determined by the parameter values jointly with the intensity of
the external noise acting on the system. However, as discussed above,
external noise may have its origin in intrinsic noise, associated
for instance to concentration fluctuations. It is a suggestive picture
that biochemical networks would have evolved in such a way that
concentration fluctuations allow the network to gain access to otherwise 
dynamically forbidden regions \cite{Vilar02}. Noise then would appear as a result 
of positive selection on gene expression levels rather than as an 
external factor causing ``defensive'' architectural evolution. To what extent evolution
can positively select for stochastic dynamics as opposed to deterministic ones is
a fundamental and intriguing question. 

The theory proposed in this paper applies to both linear and
nonlinear dependencies of reaction rates on parameters and
concentrations. In the linear case the analysis is particularly
simple. The effect of the noise is perceived by the system as a
renormalization of the relevant parameters. 

When nonlinear dependencies are considered, may be that the effect of the
noise cannot be reabsorbed as a renormalization of bare parameters only, 
and the overall dynamics is expected to be generally more complex. 
This case has been object of intense study, and the theory of
noise-induced transition has emerged \cite{Lefever84}. Because of the
nonlinearities, the effect of the noise is not limited to a shift in
the extrema of the probability distribution, to which our theory
applies for small enough noise intensities, 
but new extrema may appear, and the system can jump from one
to the other driven by random fluctuations. 
Even though this mechanism is well understood at the
kinetic level, at the systemic level its description 
requires a proper MCA, which needs to be valid for large jumps 
between stationary states. This is the subject of some recent 
literature \cite{Acerenza06,Acerenza07}, and the corresponding
analysis when extrinsic noise is present will be the subject of a
further study.

As a final remark, we would like to mention also the 
recent extensions of MCA that have been discussed in \cite{Peletier03}
to account for spatial dependencies. These are relevant for instance
in signal transduction networks, where protein diffusion plays an important role
in the signaling process \cite{Takahashi05}. Modelling these systems relies on 
defining proper Reaction-Diffusion dynamics, for which the standard MCA 
needs to be extended accordingly. Our paper follows conceptually 
the same spirit of \cite{Peletier03}, 
as it aims at identifying on general grounds 
all processes that may have relevant biological implications. 
The simultaneous treatment of both noise and spatial dependencies 
in stochastic non homogeneous systems is in turn a combination of the
theory of \cite{Peletier03} and the theory presented in this
paper. This will also be the subject of a further study.

\end{document}